\documentclass[suppldata]{interact}

\usepackage{epstopdf}% To incorporate .eps illustrations using PDFLaTeX, etc.
\usepackage[caption=false]{subfig}% Support for small, `sub' figures and tables
\usepackage{multirow}
\usepackage[natbibapa,nodoi]{apacite}
\theoremstyle{plain}% Theorem-like structures provided by amsthm.sty

\theoremstyle{definition}

\theoremstyle{remark}

\begin{document}

%\articletype{ARTICLE TEMPLATE}% Specify the article type or omit as appropriate

\title{Emotion Recognition with Forearm-based Electromyography}

\author{
\name{Muhammad Shihab Rashid\textsuperscript{a}\thanks{CONTACT Muhammad Shihab Rashid. Email: shihabrashid@iut-dhaka.edu} , Zubayet Zaman\textsuperscript{b} , Hasan Mahmud\textsuperscript{c} , Md. Kamrul Hasan\textsuperscript{d}}
\affil{Department of Computer Science and Engineering, Islamic University of Technology, Boardbazar, Gazipur, Bangladesh}
}

\maketitle

\begin{abstract}
Electromyography is an unexplored field of study when it comes to alternate input modality while interacting with a computer. However, to make computers understand human emotions is pivotal in the area of human-computer interaction and in assistive technology. Traditional input devices used currently have limitations and restrictions when it comes to express human emotions. The applications regarding computers and emotions are vast. In this paper we analyze EMG signals recorded from a low cost MyoSensor and classify them into two classes - Relaxed and Angry. In order to perform this classification we have created a dataset collected from 10 users, extracted 8 significant features and classified them using Support Vector Machine algorithm. We show uniquely that forearm-based EMG signal can express emotions. Experimental results show an accuracy of 88.1\% after 300 iterations.This shows significant opportunities in various fields of computer science such as gaming and e-learning tools where EMG signals can be used to detect human emotions and make the system provide feedback based on it. We discuss further applications of the method that seeks to expand the range of human-computer interaction beyond the button box.
\end{abstract}


\section{Introduction}
Electromyography means the recording of the electrical activity of muscle tissue, or its representation as a visual display or audible signal, using electrodes attached to the skin or inserted into the muscle. Human body is comprised of lots of nerves and sensors and they pass electric signals throughout the body. Electromyography is the study of these signals and how these impact different aspects. \\
Emotions are very crucial in our day to day lives as it can heavily influence our mood and affect our communication with others. For a long time, scientists have been trying to make computers understand our emotions. If computers can understand emotions and if they can provide feedback based on it, then the user experience while interacting with a system will be more natural. While the demand for newer and alternate method for interaction is rising high, we see methods such as body-worn computer systems, but the question of ideal input mechanism is still unanswered. Traditional input devices such as keyboard, mouse, joystick are monotonous and have restricted applications. The user experience regarding these traditional devices are very limiting. With the invent of technology we see input methods such as touch and gesture, but due to lacking of quality of cameras or algorithm accuracy, these methods cannot fully detect gestures. As a result the user experience may become cumbersome and tiring. \\
So researchers started to think of even newer ways to interact with a computer. A combination of speech and gesture seemed promising, as speech is more natural and gestures are suited for spatial or silent and harmless interaction. The applications of these methods include controlling smart glasses or watches with  subtle finger motions and gestures. But due to noise in speech data this method posed a problem. So these problems gave way to latest input mechanisms such as bio sensors which include EEG(Electroencephalography), EMG and GSR(Galvanic Skin Response). These are newer fields in the area of interaction and are still unexplored. \\
Electromyography to this day, have been used for clinical diagnosis and biomedical applications mostly\citep{technique}. %techniques of emg classify
The field of rehabilitation of motor disabilities is a key area for EMG signals. The shapes and firing rates of Motor Unit Action Potentials (MUAPs) in EMG signals provide an important source of information for the diagnosis of neuromuscular disorders. So far research and extensive efforts have been made in the area to develop better algorithms, upgrading existing technologies, improving detecting mechanism and to acquire accurate data. Another application of EMG includes in the field of Robotics where artificial hand control and movement, grasp control etc. However, EMG is yet to be used as an input modality in interaction and our paper proposes that as EMG can convey emotions, the machine can give accurate responses. \\
We believe the process is more natural when we try to convey emotions using muscles. As it is dependent on our reflexes, when we are in an emotional state, our muscles automatically give responses based on a certain emotion. For which the responses can be more or less accurately recorded by sensors. We thought the responses might follow a pattern in case of all the individuals so we did the experiment and got promising results. \\
The potential application of our method includes emotion-based game difficulties. Sometimes the gamers feel stressed while playing games and this may cause health difficulties and may change the mood. EMG can be used to detect user stress level or emotion status while playing the game and based on the status, the computer can change the difficulty level. For example, when the user gets stressed, the game level may offer an extra booster so the player can control his or her stress level. Our method can also be integrated with virtual reality technologies where user emotion adds an extra layer to the user experience.

\section{Related Works}
There are different classification systems for emotion. According to Plutchik\citep{plutchik} emotion can be of eight types. Anger, fear, sadness, disgust, surprise, anticipation, acceptance and joy. All other emotions can be formed by mixing these basic emotions. %R. Plutchik, Emotions and life : perspectives from
%psychology, biology, and evolution, 1st ed. Washington, DC:
%American Psychological Association, 2003.
The most widely used system is Russel's bipolar system where emotion is classified into two categories: Arousal and Valence\citep{bipolar}. %J. A. Russell, “Affective space is bipolar,” Journal of
%Personality and Social Psychology, vol. 37, 1979, pp. 345-
%356.
Arousal ranges from sad/angry to excited/joyful and valence ranges from negative to positive. In this paper, we have chosen two contrasting emotions such as Angry and Relaxed so that we can differentiate between the emotions. The second reason being, in the applications of EMG such as in gaming, people usually express these two types of emotions.
\begin{figure}[!ht]
	\hfill \includegraphics[width=0.7\linewidth]{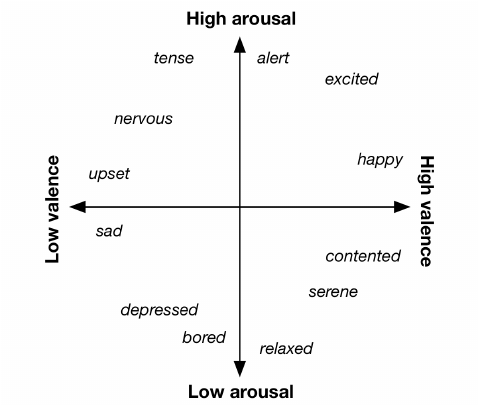} \hspace*{\fill}
	\caption{Valence Arousal Matrix}
	\label{fig}
\end{figure}
\\
In the Figure 1, we can see the two dimensional valence arousal matrix. Where emotions such as Happy or excited are regarded as high valence emotions and being upset or sad is regarded as low valence emotion. High arousal emotions are alert and tense and low arousal emotions are relaxed. \\
Although detection of human emotions in the area of human-computer interaction is a newer methodology itself, there are some significant research done in this area. However, whether emotion detection can be done through electromyography data is still to be answered. Similar mechanisms have been used in detecting emotions in recent years.  \\
Firstly, as per human anatomy, brain is the most fundamental source of emotion and best way to measure the neurological changes. This type of process is called Electroencephalography or EEG. In recent years MRI also shows great promise in detecting human emotion. In \citep{realeeg}\citep{noisy} %real time eeg based human
authors show that using a brain signal sensor, emotions can be detected in the valence-arousal two dimensional model. But the author proposes some challenges in working with brain sensors. The sensors are noisy as our brain handles different functions at a certain point in time. So the separation of noise with actual data is quite difficult. \\
Secondly, facial muscles or expressions plays a vital role for both verbal and nonverbal communication. Due to changes in mental state, users facial expressions also change, but these expressions are somewhat forced, it does not happen naturally. The added challenge of analysis and processing of huge data with complex algorithms to generate probable emotion is difficult to implement in smaller hardware or devices. So a more convenient way is necessary. \\
Third, audio signals or voice can also play a part. Users vocal chords can carry significant variations or emotional information. From the pitch, intensity and pitch contours are estimated. Speech data most of the times do not correlate with the meaning the user intended to say. Most times it is misinterpreted by computer. \\
Newer ways of emotion recognition such as haptile feedback \citep{haptic} %haptic
which is based on users touch input seem promising\citep{ux}\citep{tactile}. However, users touch contain very little information regarding emotional activity which proposes as a problem. In \citep{rasam} %Rassam
the authors propose that emotions can be expressed depending on users sitting position and also using smartphones different sensors. These methods are dependent on the user having a smartphone or sitting at a certain position. Fusing all these methods together creates a multimodal emotion recognition system. The multimodal approach is a lot more dynamic but it has to correlate multiple types of inputs and finally integrate with probabilistic models. \\
In \citep{noninvasive}\citep{bio} they used a combination of biological sensors such as EEG, EMG and GSR or biological gloves to detect emotions. Their study does not portray the possibility of detecting emotion just based on EMG data.  \\
Regarding applications of this research domain, in \citep{rehab} they have built a device also using myosensors to interact with a computer using facial muscles and facial mimicry. But these approach has limiting nature because it only uses facial muscles where the muscle movements are strict.
\subsection{Feature Extraction, Selection}
Even though forearm-based EMG signals have not been used to classify emotions but EMG signals in general have been used in different researchs. In \citep{surfacewavelet} they have extracted features using wavelet transform. In \citep{technique}, there are around 50 types of features described in the time domain and the frequency domain. In \citep{featuretime}, the authors tried to extract features based on the biceps muscle. The have extracted the following three features: Maximum amplitude, Standard Deviation and Root Mean Square.
\subsection{Studies Regarding EMG}
Although forearm-based Electromyography has not been used to detect emotion or been used in human computer interaction applications, there has been studies regarding EMG. In \citep{enable} they have
shown how forearm EMG can be used to
detect and decode human muscular movement in real time, thus enabling interactive finger gesture interaction. They have also explored techniques that will enable people to interact with computers when their hands are unencumbered but a hand-held device is impractical. \\
In \citep{myopoint} they have developed a mid-air, free hand pointing and clicking interaction mechanism using electromyographic (EMG) and inertial measurement unit (IMU) input from a consumer armband device. They have shown significant improvement when clicking and pointing with Myo devices. Also, this indicates, myo devices can be a modality of interaction with computers. Although they have shown this in a limited scale, it is an indicator that there are vast possibilities of EMG being used as an interaction modality. \\
In \citep{feasible} they have shown, using a Muscle Computer Interface (muCIs), gestures can be classified using an off-the-shelf electromyography device. They were able to differentiate position and pressure of finger presses, as well as recognizing tapping and gestures of lifting across all the fingers of one hand. In \citep{advance} they used a sensor array based muscle-computer interface for gesture recognition. These studies show that, EMG can be a viable interaction method.
\section{Proposed Approach}
We propose an emotion recognition system through EMG signals. The architectural diagram is given in Figure 2. In general, the proposed system consists of the following steps:
\begin{itemize}
    \item Data collection
    \item Pre-Processing
    \item Feature extraction and selection
    \item Training set generation
    \item Classification using SVM
\end{itemize}
 \begin{figure}[htbp]
 	\includegraphics[width=\linewidth]{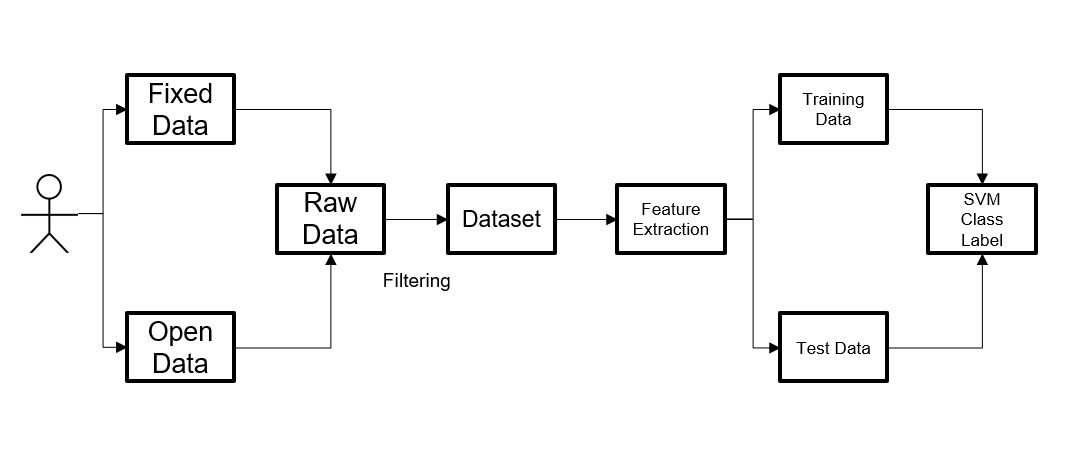}
 	\caption{Architectural Diagram of Proposed Approach}
 	\label{fig}
 \end{figure}

\subsection{Experiment Setup}
For conducting the experiment, we built a wearable device using low cost MyoWare Muscle sensor by Advancer Technologies and an Arduino Mega board. The muscle sensor device has three bio-sensor electrodes of which one is used as a reference electrode. The device can be worn on any body part that contains muscle, but placing it on the forearm gives the most accurate results.
A simple AnalogReadSignal function was used using Arduino software to capture the sensor readings.
\begin{figure}[htbp]
		\hfill \includegraphics[width=0.5\linewidth]{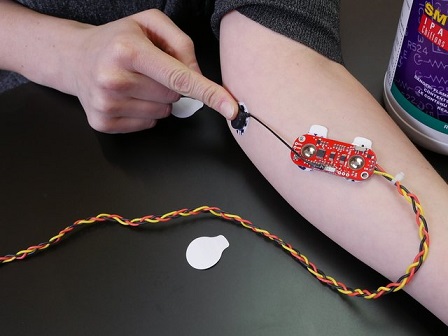} \hspace*{\fill}
	\caption{MyoWare Muscle Sensor}
	\label{fig}
\end{figure}
In Figure 2 we can see the myoware device put on the forearm. The wire is connected to the Arduino mega board. The device is very small in size so it has better mobility than other traditional devices.
\subsection{Data Collection}
We have selected 10 university going individuals consisting of both male and female as our users. The wearable device was placed on the forearm of the users dominating hand. The experiment followed two steps: \\
Firstly, in \textbf{"Fixed"} typing method, the users were told to type a certain paragraph exactly as it is in a \textbf{"Relaxed"} state. The users were shown pictures that induce relaxing features of ones mind. They listened to soothing music for about one minute.  On the computer screen a paragraph was displayed taken from the famous children's novel \textit{Alice's adventures in wonderland}(Carroll 2008)According to Epp et al.(2011) \citep{keystroke}
\begin{figure}[htbp]
			\hfill \includegraphics[width=0.7\linewidth]{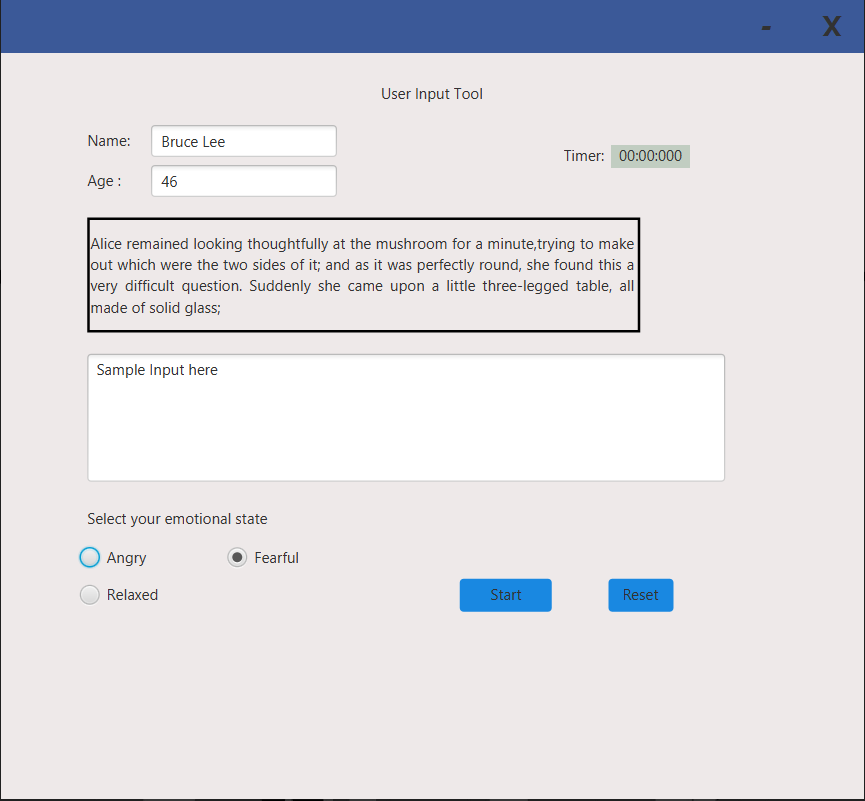} \hspace*{\fill}
	\caption{User Input Tool}
	\label{fig}
\end{figure}
The reasons behind
choosing these text excerpts are that they have relatively simple sentence structure, an absence of large uncommon words and that each piece of text is roughly the same length. The raw data from the sensor along with the timing was recorded. The users were again told to type the passage but this time in a \textbf{"Angry"} emotional state. They were shown anger inducing picture on the computer screen. \\
In the \textbf{"Open"} typing stage, the users were told to type anything they want for one minute in Relaxed and Angry emotional state consecutively. Again the raw sensor readings were recorded. \\
The reason for choosing this type of experimental setup is that, while typing, the muscles of forearm and fingers are most active and the responses can be accurately collected. \\
Figure 3 shows the input interface tool that was used to conduct the experiment.
\subsection{Raw EMG Data}
The sensor readings were collected from the Arduino software and saved it into excel using the terraterm software. The sensor readings are integer values ranging from 0 to 999. \\
\begin{figure}[htbp]
			\hfill \includegraphics[width=0.8\linewidth]{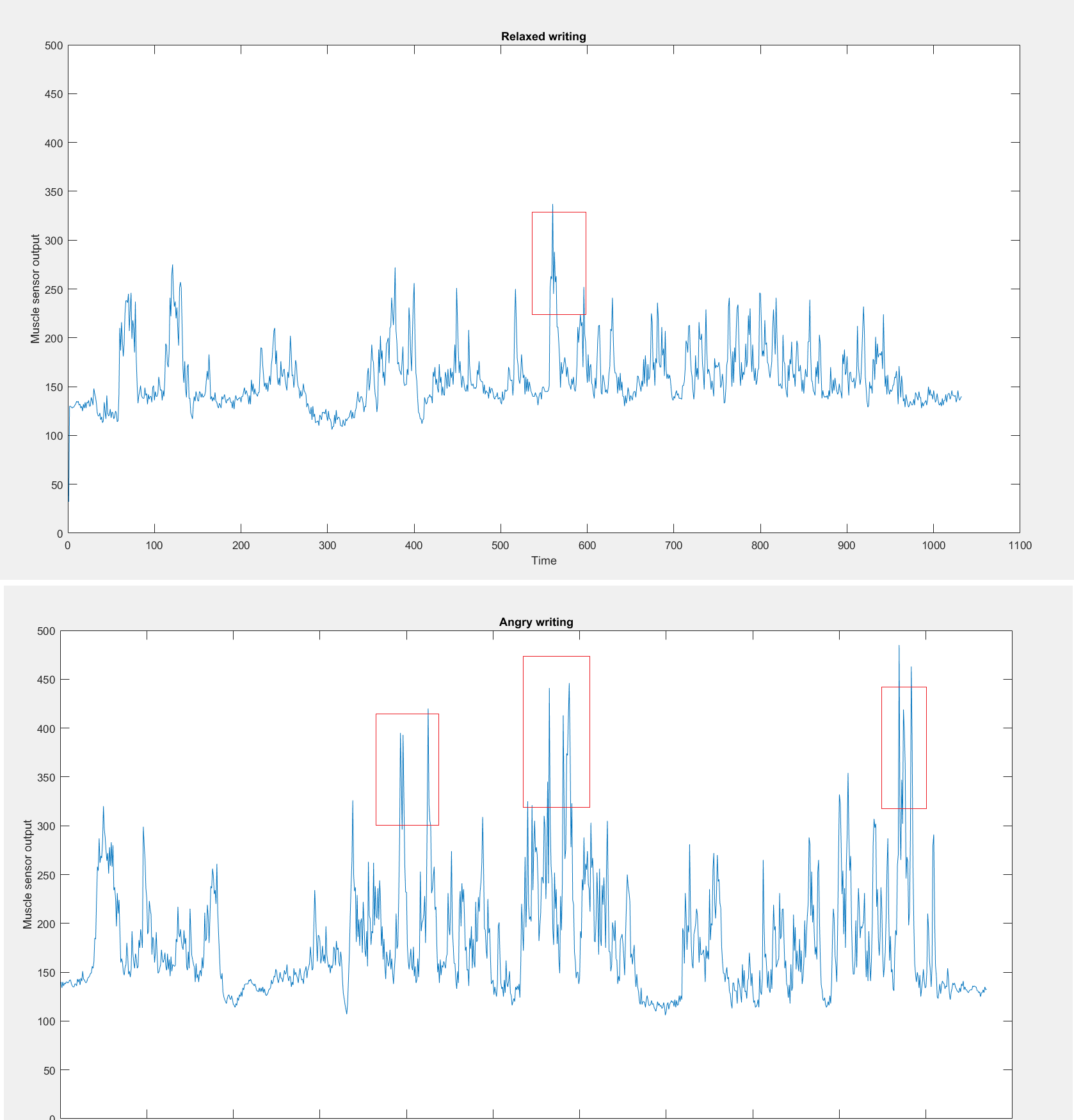} \hspace*{\fill}
	\caption{Raw EMG Data Showing Relaxed and Angry state}
	\label{fig}
\end{figure}
We can see from Figure 4 that, in 'relaxed' data it follows a simple pattern with very few spikes (highlighted in red). The signal is kept at a normal level and it shows that human muscles stay at a stable pace throughout.
%\begin{figure}[htbp]
%	\includegraphics[width=\linewidth]{angry}
%	\caption{Raw Angry EMG Data}
%	\label{fig}
%\end{figure}
In the raw angry emg data from the user, it is apparent from the naked eye that the pattern is not similar. It has more spikes and ups and downs from the relaxed one. So we can guess that if this data is fed into the machine to report for a pattern, we may find it.
\subsection{Filtering}
The MyoWare sensor has some built in filtering mechanisms to reduce noise which makes it the ideal device to capture muscle data. In our experiment, we told the users to not type anything for the first 10 seconds and the last 5 seconds after they have done typing the paragraph. Our initial filtering process includes the reduction of those irrelevant data from the raw ones.
\subsection{Feature Extraction}
To perform any machine learning approach, we have to generate features first. Features are characteristics of any dataset. But different types of features can be taken, from that which ones we consider is the key thing to do. From papers regarding feature extraction from EMG or EEG signal, it is understood that features are divided into two domains: Time domain and Frequency domain \citep{technique}. \\
But frequency domain features create redundancy while doing classification as we see in \citep{feature}. So we have selected the following 8 features form time domain: \\
%lots of citations needed, check the original paper 
\begin{itemize}
	\item \textbf{Maximum Peak in a Timespan:} This means in a specific timespan what is the maximum peak of raw data
	\item \textbf{Mean Absolute Value:} Mean absolute value (MAV) is one of the most popular used in
	EMG signal analysis. There are many given names for calling this feature; for instance, average rectified value
	(ARV), averaged absolute value (AAV), integral of absolute value
	(IAV), and the first order of v-Order features (V1). MAV feature is
	an average of absolute value of the EMG signal amplitude in a segment
	\item \textbf{Mean Absolute Value Slope:} Mean absolute value slope (MAVSLP) is a modified version of
	MAV feature to establish multiple features. Differences between MAVs of the adjacent segments
	are determined. 
	\item \textbf{Picks Above Average Frequency:} This means how many peaks are there above the average frequency
	\item \textbf{Root Mean Square:} Root mean square (RMS) is another popular feature in analysis	of the EMG signal. It is modeled as amplitude modulated Gaussian random process
	whose relates to constant force and non-fatiguing contraction. It
	is also similar to standard deviation method
	\item \textbf{Average Amplitude Change:} Average of how many times amplitudes change in a timespan signal
	\item \textbf{Difference Absolute Standard Deviation Value(DASDV):} Difference absolute standard deviation value (DASDV) is look
	like RMS feature, in other words, it is a standard deviation value
	of the wavelength
	\item \textbf{Waveform Length:} Waveform length (WL) is a measure of complexity of the EMG
	signal. It is defined
	as cumulative length of the EMG waveform over the time segment.
	Some literatures called this feature as wavelength (WAVE).
\end{itemize}
Among these features some are selected from \citep{feature} and some features such as Maximum Peak in a Timespan, Picks Above Average Frequency, Avg Amplitude Change are selected from our own intuition. We believe these features are closely related with electromyography signal deviations when users move from one emotional state to another.
\subsection{Training Data}
We have in total 10 users, from each user we have collected 4 types of data. They are: Fixed Relaxed (Relaxed data collected in fixed format), Fixed Angry, Open Relaxed, Open Angry. There are 40 types of data or 40 "rows" in the training data matrix. Each data is divided into 10 time slots. In each time slots there are 8 features. So there are 80 features in total per row. In total the matrix size is 40x80. The process is described in the Figure 6  \\
\begin{figure}[htbp]
	\includegraphics[width=\linewidth]{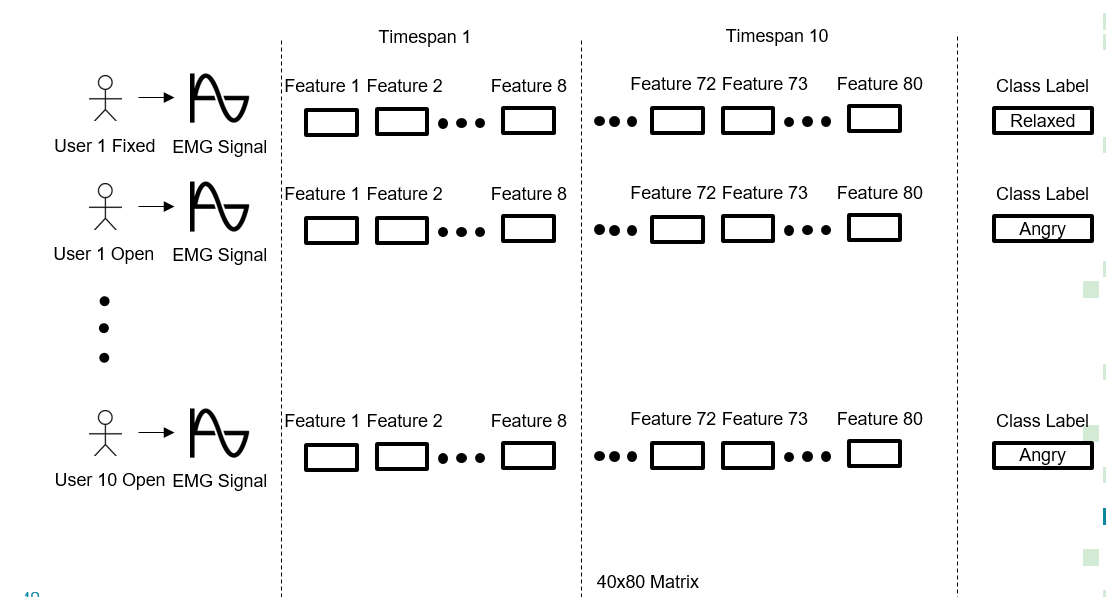}
	\caption{Feature Collection Process}
	\label{fig}
\end{figure}
The matrix(Figure 7) looks like the following in MatLab: \\

\begin{figure}[htbp]
	\includegraphics[width=\linewidth]{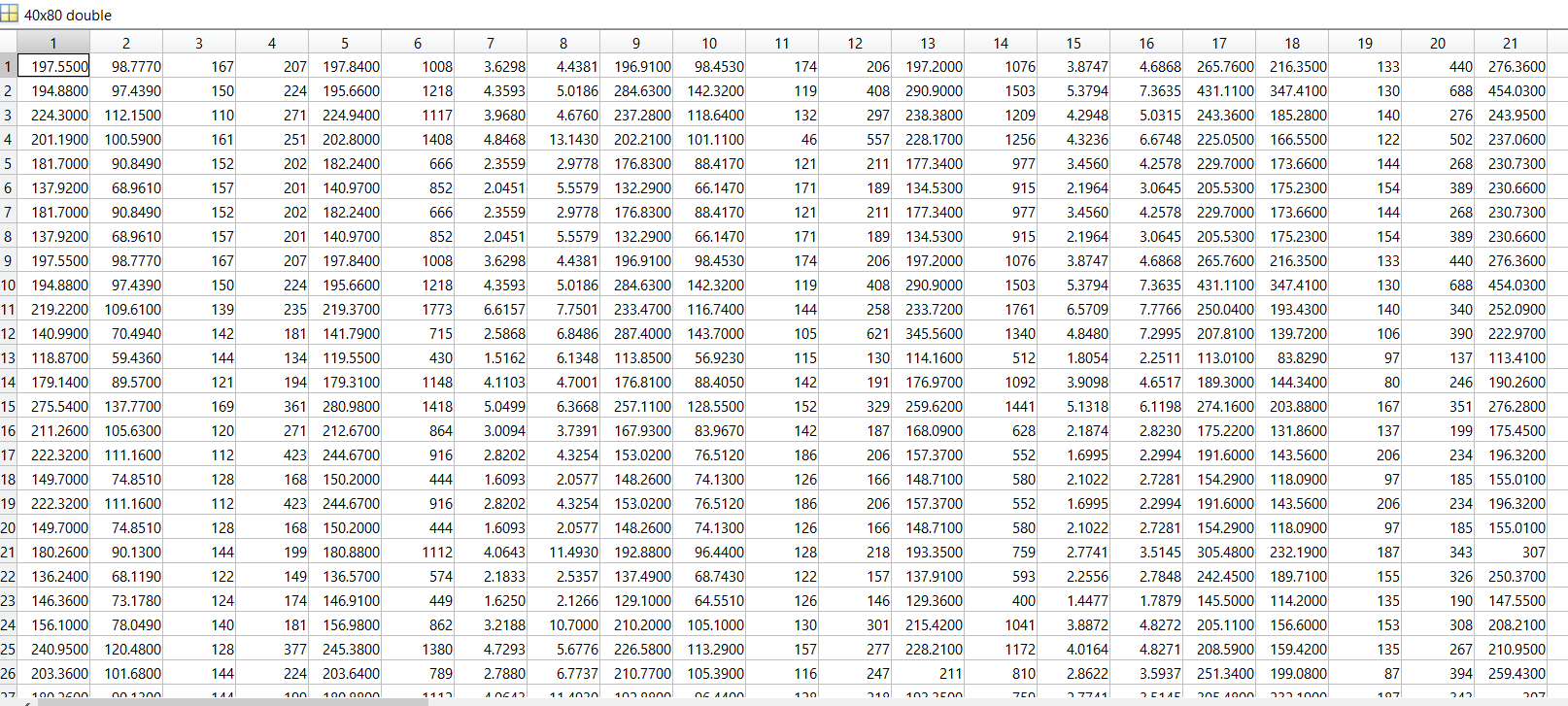}
	\caption{Training Data Set}
	\label{fig}
\end{figure}
\subsection{Selecting Best Features}
We see from Figure 7 that, there are 80 features in total per user data. This is a large number which may cause over-fitting of data. So, we made an algorithm that selects the best 5 features that gives the highest accuracy for that specific training data. The algorithm automatically takes 5 feature sets and then calculates. It performs all the combinations and sees which set produces the highest accuracy. Those 5 features are selected for classifying test data. The reason why 5 features are selected is that, it gives the best accuracy overall. We have performed the algorithm using 3,7,9 features but we see that 5 is the minimum number that gives the best overall accuracy.
\subsection{Classification}
We have used two approaches to generate test and train data: \\
\begin{itemize}
	\item \textbf{Leave One Out:} Here out of 10 users, 1 user was kept aside for test data and the rest 9 were used for train.
	\item \textbf{80-20 Method:} In this approach, 20\% of data is selected as random for testing. The rest are training data 
\end{itemize}
In our research we have followed both the techniques.
\subsection{Recognition Using SVM}
We have used Support Vector Machine with 5 folds cross validation to classify our dataset. Our class labels are "Relaxed" and "Angry". The kernel used is called "Radial Basis Function. The kernel function is a measure of similarity between two sets of features. The SVM algorithm picks up the best 5 features from the 80 feature and matches the features. \\
\begin{figure}[htbp]
		\hfill \includegraphics[width=0.5\linewidth]{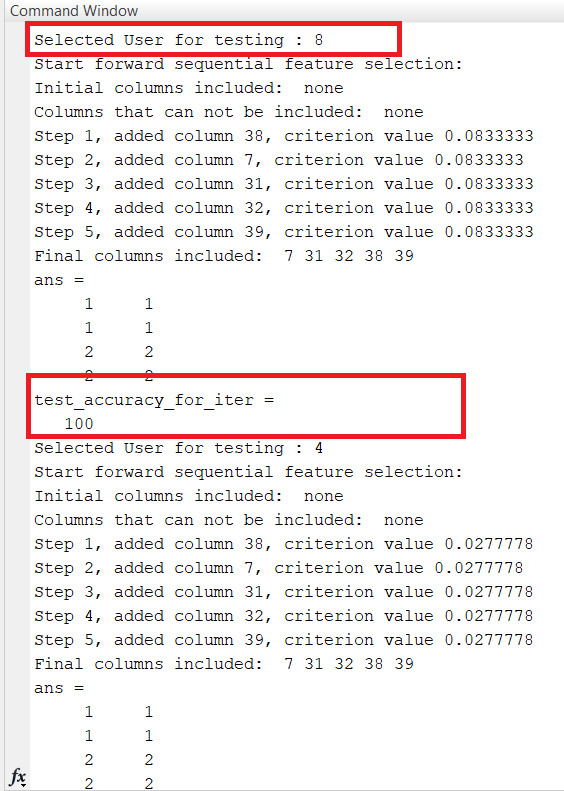} \hspace*{\fill}
	\caption{Leave One Out Method}
	\label{fig}
\end{figure}
In Figure 8, we see the output of the algorithm in matlab for the leave one out method, where the accuracy of that particular iteration was 100\%. In total 400 iterations were done and the average accuracy after 400 iterations were 93\%. 
\begin{figure}[!ht]
		\hfill \includegraphics[width=0.5\linewidth]{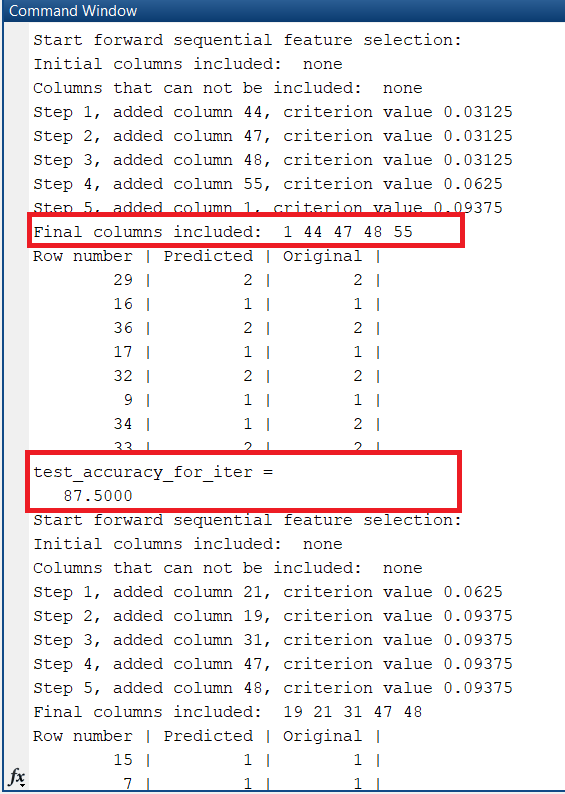} \hspace*{\fill}
	\caption{Twenty Eighty Method}
	\label{fig}
\end{figure}
In Figure 9 we see output of the algorithm for the twenty eighty method. The accuracy for that specific iteration was 87.5\%. There were in total 400 iterations and the average accuracy was 88\%.
\subsection{Accuracy}
Following the Leave One Out approach we got around 93\% accuracy for 400 iterations. And following 80-20 approach we got around 88\% accuracy for 400 iterations.
\subsection{Confusion Matrix}
The confusion matrix for the leave one out method is showed in Table 1.: \\
\begin{table}[!ht]
		\tbl{Confusion Matrix For Leave One Out Method}
{
	\begin{tabular}{|c|c|c|c|}
		\hline
		\multicolumn{4}{|c|}{Original}                         \\ \hline
		\multirow{3}{*}{Predicted} &         & Angry & Relaxed \\ \cline{2-4} 
		& Angry   & 777   & 88      \\ \cline{2-4} 
		& Relaxed & 23    & 712     \\ \hline
	\end{tabular}}
\label{my-label2}
\end{table}
From the confusion matrix(Figure 10) we can deduct a few things. The number of misclassification is less when in case of mistaking angry data for relax data. This is a positive point for our approach. Because this means if the user is angry, in very few cases it will be mistaken for him/her being relaxed. In our application areas for example, when a driver is driving a car, if he gets angry but our system deducts it as a relax emotional state then that creates a problem. The system will allow the driver to continue driving and risk of accident will increase. So it is a good thing that our system has few numbers of misclassification.
The confusion matrix for the 80/20 method is shown in Table 2: \\
\begin{table}[!ht]
		\tbl{Confusion Matrix For 80-20 Method}
{
	\begin{tabular}{|c|c|c|c|}
		\hline
		\multicolumn{4}{|c|}{Original}                         \\ \hline
		\multirow{3}{*}{Predicted} &         & Angry & Relaxed \\ \cline{2-4} 
		& Angry   & 1473  & 331     \\ \cline{2-4} 
		& Relaxed & 76    & 1320    \\ \hline
	\end{tabular}}
\label{my-label3}
\end{table}
In the 80-20 method(Figure 11) we can also see that the misclassification number is very less when mistaking angry data for relax data.\\
From the confusion matrix we can calculate the following data showed in the Table 3:\\
\begin{table}[!ht]
\tbl{Confusion Matrix Statistics}
{		\begin{tabular}{|l|l|l|} 
			\hline
			\textbf{Measure}             & \textbf{Value}  & \textbf{Derivations}                \\ \hline
			Accuracy            & 0.9306 & ACC = (TP + TN) / (P + N)  \\ \hline 
			Precision           & 0.8983 & PPV = TP / (TP + FP)       \\ \hline 
			Sensitivity         & 0.9713 & TPR = TP / (TP + FN)       \\ \hline 
			Specificity         & 0.8900 & SPC = TN / (FP + TN)       \\ \hline 
			False Positive Rate & 0.1100 & FPR = FP / (FP + TN)       \\ \hline 
			False Negative Rate & 0.0288 & FNR = FN / (FN + TP)       \\ \hline 
			F1 Score            & 0.9333 & F1 = 2TP / (2TP + FP + FN) \\ \hline 
		\end{tabular}}
	\label{my-label}
\end{table}
The notations used in the table are TP: True Positive, TN: True Negative, P: Total Positive, N: Total Negative, FP: False Positive, FN: False Negative.
\section{Conclusions and Future Works}
In this paper, we tried to explore a field of interaction that has yet not been explored. We have tried to find out whether EMG could be a potential method to interact with a system. The results we found were satisfactory but it also has some limitations. The sensor that we used was a low cost one and their were some noise in the raw data.  The implications of our research can be vast. As we have established that emotions are significantly related with our muscles, we can argue that we can give this emotional state as an input to computer systems and the system will be able to provide feedback based on users emotional state. \\

For our future studies we are planning to build a wireless version of our device. Currently the device we are using is wired and it sometimes create problems when moving arm. So if the device is wireless then it will be easy for the users to move their arm freely. Furthermore we are planning to improve our accuracy by using different machine learning algorithms. Next we are planning to build an interactive virtual table tennis application to conduct an experiment. The virtual table tennis application will be played by forearm movement where there will be our device set up on the arm. The users will be able to provide "power" as a feedback to the game. The more the user grips the bat, the more the projectile of the ball will be. We will let our users play both versions of the game and let them decide which method they prefer.

\bibliographystyle{apacite}
\bibliography{paper}

\end{document}